\documentclass[twocolumn,aps,pra,superscriptaddress,floatfix]{revtex4}
\usepackage{graphicx}
\usepackage{times}
\usepackage{nicefrac}
\usepackage{amsmath}
\usepackage{amsfonts}
\usepackage{amssymb}
\usepackage{amsthm}
\usepackage{epsf}
\usepackage{bm}
\usepackage{bbm}
\usepackage{times}

\usepackage{dcolumn}

\newcolumntype{.}{D{x}{}{-1}}

\begin{document}

\newcommand{\half}{\frac12}
\newcommand{\vare}{\varepsilon}
\newcommand{\eps}{\epsilon}
\newcommand{\pr}{^{\prime}}
\newcommand{\ppr}{^{\prime\prime}}
\newcommand{\pp}{{p^{\prime}}}
\newcommand{\ppp}{{p^{\prime\prime}}}
\newcommand{\hp}{\hat{\bfp}}
\newcommand{\hr}{\hat{\bfr}}
\newcommand{\hk}{\hat{\bfk}}
\newcommand{\hx}{\hat{\bfx}}
\newcommand{\hpp}{\hat{\bfpp}}
\newcommand{\hq}{\hat{\bfq}}
\newcommand{\rqq}{{\rm q}}
\newcommand{\bfk}{{\bm{k}}}
\newcommand{\bfp}{{\bm{p}}}
\newcommand{\bfq}{{\bm{q}}}
\newcommand{\bfr}{{\bm{r}}}
\newcommand{\bfx}{{\bm{x}}}
\newcommand{\bfy}{{\bm{y}}}
\newcommand{\bfz}{{\bm{z}}}
\newcommand{\bfpp}{{\bm{\pp}}}
\newcommand{\bfppp}{{\bm{\ppp}}}
\newcommand{\balpha}{\bm{\alpha}}
\newcommand{\bvare}{\bm{\vare}}
\newcommand{\bgamma}{\bm{\gamma}}
\newcommand{\bGamma}{\bm{\Gamma}}
\newcommand{\bLambda}{\bm{\Lambda}}
\newcommand{\bmu}{\bm{\mu}}
\newcommand{\bnabla}{\bm{\nabla}}
\newcommand{\bvarrho}{\bm{\varrho}}
\newcommand{\bsigma}{\bm{\sigma}}
\newcommand{\bTheta}{\bm{\Theta}}
\newcommand{\bphi}{\bm{\phi}}
\newcommand{\bomega}{\bm{\omega}}
\newcommand{\intzo}{\int_0^1}
\newcommand{\intinf}{\int^{\infty}_{-\infty}}
\newcommand{\lbr}{\langle}
\newcommand{\rbr}{\rangle}
\newcommand{\ThreeJ}[6]{
        \left(
        \begin{array}{ccc}
        #1  & #2  & #3 \\
        #4  & #5  & #6 \\
        \end{array}
        \right)
        }
\newcommand{\SixJ}[6]{
        \left\{
        \begin{array}{ccc}
        #1  & #2  & #3 \\
        #4  & #5  & #6 \\
        \end{array}
        \right\}
        }
\newcommand{\NineJ}[9]{
        \left\{
        \begin{array}{ccc}
        #1  & #2  & #3 \\
        #4  & #5  & #6 \\
        #7  & #8  & #9 \\
        \end{array}
        \right\}
        }
\newcommand{\Vector}[2]{
        \left(
        \begin{array}{c}
        #1     \\
        #2     \\
        \end{array}
        \right)
        }

\newcommand{\Dmatrix}[4]{
        \left(
        \begin{array}{cc}
        #1  & #2   \\
        #3  & #4   \\
        \end{array}
        \right)
        }
\newcommand{\Dcase}[4]{
        \left\{
        \begin{array}{cl}
        #1  & #2   \\
        #3  & #4   \\
        \end{array}
        \right.
        }
\newcommand{\cross}[1]{#1\!\!\!/}

\newcommand{\Za}{{Z \alpha}}
\newcommand{\im}{{ i}}

\newcommand{\lapprox }{{\lower0.8ex\hbox{$\buildrel <\over\sim$}}}

\def\ket#1{ $ \left\vert  #1   \right\rangle $ }
\def\ketm#1{  \left\vert  #1   \right\rangle   }
\def\bra#1{ $ \left\langle  #1   \right\vert $ }
\def\bram#1{  \left\langle  #1   \right\vert   }
\def\spr#1#2{ $ \left\langle #1 \left\vert \right. #2 \right\rangle $ }
\def\sprm#1#2{  \left\langle #1 \left\vert \right. #2 \right\rangle   }
\def\me#1#2#3{ $ \left\langle #1 \left\vert  #2 \right\vert #3 \right\rangle $ }
\def\mem#1#2#3{  \left\langle #1 \left\vert  #2 \right\vert #3 \right\rangle   }
\def\redme#1#2#3{ $ \left\langle #1 \left\Vert
                  #2 \right\Vert #3 \right\rangle $ }
\def\redmem#1#2#3{  \left\langle #1 \left\Vert
                  #2 \right\Vert #3 \right\rangle   }
\def\rmem#1#2#3{  \left\langle #1 \left\vert \left\vert  #2
                  \right\vert \right\vert #3 \right\rangle   }

\title{Diagnostics of polarization purity of x-rays by means of Rayleigh scattering}

\author{A.~Surzhykov}
\affiliation{Physikalisch-Technische Bundesanstalt, D-38116 Braunschweig, Germany}
\affiliation{Technische Universit\"at Braunschweig, D-38106 Braunschweig, Germany}

\author{V.~A. Yerokhin}
\affiliation{Physikalisch-Technische Bundesanstalt, D-38116 Braunschweig, Germany}
\affiliation{Center for Advanced Studies, Peter the Great St.~Petersburg Polytechnic University,
195251 St.~Petersburg, Russia}

\author{S.~Fritzsche}
\address{Helmholtz-Institut Jena, D--07743 Jena, Germany}
\address{Theoretisch-Physikalisches Institut, Friedrich--Schiller-Universit\"at Jena,
D-07743 Jena, Germany}

\author{A.~V.~Volotka}
\address{Helmholtz-Institut Jena, D--07743 Jena, Germany}
\address{Department of Physics, St.~Petersburg State University, 198504 St.~Petersburg, Russia}

\begin{abstract}

The synchrotron radiation is commonly known to be completely linearly polarized when observed in
the orbital plane of the synchrotron motion. Under actual experimental conditions, however, the
degree of polarization of the synchrotron radiation may be lower than the ideal 100\%. We
demonstrate that even tiny impurities of polarization of the incident radiation can drastically
affect the polarization of the elastically scattered light. We propose to use this effect as a
precision tool for the diagnostics of the polarization purity of the synchrotron radiation. Two
variants of the diagnostics method are proposed. The first one is based on the polarization
measurements of the scattered radiation and relies on theoretical calculations of the transition
amplitudes. The second one involves simultaneous measurements of the polarization and the cross
sections of the scattered radiation and is independent of theoretical amplitudes.

\end{abstract}

\maketitle

\section{Introduction}

Advent of third-generation synchrotron radiation facilities made possible scattering experiments of
a novel type, where the linear polarizations of both the incident and the scattered particles
(photons and/or electrons) were detected \cite{tashenov:11,maertin:12,tashenov:13,kovtun:15}. These
experiments measured polarization correlations and the polarization transfer between the incident
and scattered beams. They demonstrated that the detection of polarization properties of the
scattered particles could be used for an accurate determination of the polarization properties of
the incident beam.

In the present work we consider the elastic (Rayleigh) scattering of the synchrotron radiation off
a closed-shell atomic target. The synchrotron radiation is known to be 100\% linearly polarized
when observed in the orbital plane of the synchrotron motion. It is also well-known that the
elastic scattering of a fully linearly polarized x-rays on a closed-shell atomic target preserves
the polarization and thus yields a fully linearly polarized scattered radiation. This is an exact
statement that follows from the symmetry requirements \cite{roy:86}. As a consequence, one might
expect a fully linearly polarized scattered radiation as an outcome of the process.

A closer examination, however, shows that this is not always the case. In particular, a recent
experiment \cite{blumenhagen:16} found a strong depolarization of the synchrotron radiation
scattered off the gold target, with the degree of polarization of $27\pm 12\%$ if observed at the
angle $\theta = 90^{\circ}$. The explanation of this phenomenon is twofold: first, in practice, the
synchrotron radiation turned out to be slightly depolarized (in the case of
Ref.~\cite{blumenhagen:16}, by about 2\%) and, second, a small depolarization of the initial
radiation had a drastically amplified effect on the polarization of the scattered radiation.

In this work, we present a detailed investigation of the effect of strong depolarization of the
elastically scattered light. We demonstrate that this effect can be used as a precision tool for
the diagnostics of the polarization purity of the x-ray radiation. The proposed method can detect
very small deviations of the degree of the linear polarization from 100\%, which is beyond
detection possibilities of other methods such as the traditional Compton polarimetry.

The proposed detection technique can become an important tool in investigations of various effects
that use the synchrotron radiation and strongly depend on purity of its polarization, such as
studies of the x-ray magnetic linear/circular dichroism \cite{stohr:98,ghidini:15} and the x-ray
magnetic scattering \cite{durr:99}, nuclear scattering diagnostics methods \cite{arthur:96}, etc.

\section{Basic theory}
\label{sec:basic}

We consider the process of the Raleigh scattering, namely, the elastic scattering of a photon on
bound atomic electrons. Initially, we have an incoming photon with the momentum ${\bm k}_i$ and
polarization ${\bm \epsilon}_i$ and an atomic state $|i\big>$ with the energy $E_i$, the total
angular momentum $J_i$ and its projection $M_i$. After the scattering, we have the scattered photon
momentum ${\bm k}_f$ ($|{\bm k}_f| = |{\bm k}_i| = \omega$) and polarization ${\bm \epsilon}_f$ and
an atomic state $|f\big>$ with the energy $E_f = E_i$, the total angular momentum $J_f = J_i$ and
its projection $M_f$. The amplitude of the process is given by \cite{roy:99}
\begin{align}
   \label{eq1}
   {\cal M}_{fi} = &
 \,-r_0\,mc^2\,
   \sum\!\!\!\!\!\!\!\!\int\limits_{\nu}
\Bigg[ \frac{\mem{f}{\hat{\mathcal{R}}^\dag({\bm k}_f, {\bm \epsilon}_f)}{\nu} \: \mem{\nu}{\hat{\mathcal{R}}({\bm k}_i, {\bm \epsilon}_i)}{i}}
      {E_i - E_{\nu} + \omega + i0}
   \nonumber \\
   & + \frac{\mem{f}{\hat{\mathcal{R}}({\bm k}_i, {\bm \epsilon}_i)}{\nu} \:
   \mem{\nu}{\hat{\mathcal{R}}^\dag({\bm k}_f, {\bm \epsilon}_f)}{i}}{E_i - E_{\nu} - \omega - i0}
   \Bigg] \, ,
\end{align}
where $r_0 = e^2mc^2$ is the classical electron radius, the summation over $\nu$ goes over all
possible intermediate electronic states of the atom, and $\hat{\mathcal{R}}$
$(\hat{\mathcal{R}}^\dag)$ is the operator of the absorbtion (emission) of a photon in the Coulomb
(velocity) gauge,
\begin{equation}
   \hat{\mathcal{R}}= \sum_m {\bm \epsilon} \cdot{\bm \alpha}_m \,  {\rm e}^{i {\bm k} \cdot {\bm r}_m} \,.
\end{equation}
Here, $m$ numerates electrons in the atom, ${\bm \epsilon}$ is the polarization vector of the
photon and ${\bm \alpha}$ is the vector of Dirac $\alpha$ matrices. The amplitude ${\cal M}_{fi}$
was considered in detail in the literature \cite{johnson:76,kissel:80,kane:86} as well as in our
previous studies \cite{surzhykov:13,surzhykov:15} and need not be discussed further.

The angle-differential cross section of the process is connected to the amplitude by
\begin{equation}
\frac{d\sigma}{d\Omega} = \big|{\cal M}_{fi}\big|^2\,.
\end{equation}
It depends on the polarization vectors of the initial and the scattered photons. In order to obtain
the angle-differential cross section observed in an experiment, the above expression should be
averaged over the polarization distribution of the incident radiation and, if the final
polarization is not detected, summed over the polarizations of the scattered radiation.

In the present work we are interested in the scattering off the closed-shell atoms, for which $J_i
= M_i = J_f = M_f = 0$. In this case, due to symmetry considerations, the transition amplitude can
be parameterized \cite{roy:86} by the parallel ${\cal A}_{||}$ and perpendicular ${\cal A}_{\perp}$
amplitudes,
\begin{equation} \label{eq2}
{\cal M}_{fi} = {\bm \epsilon}_{i||}\cdot{\bm \epsilon}_{f||}^*\,{\cal A}_{||}
  + {\bm \epsilon}_{i\perp}\cdot{\bm \epsilon}_{f\perp}^*\,{\cal A}_{\perp}\,,
\end{equation}
where ${\bm \epsilon}_{||}$ and ${\bm \epsilon}_{\perp}$ are the components of the polarization
vector parallel and perpendicular to the scattering plane (i.e., the plane spanned by vectors ${\bm
k}_i$ and vectors ${\bm k}_f$). The amplitudes ${\cal A}_{||}$ and ${\cal A}_{\perp}$ depend only
the photon energy $\omega$ and the scattering angle $\theta$ (i.e., the angle between ${\bm k}_i$
and ${\bm k}_f$).

We can use Eq.~(\ref{eq2}) to express all observable quantities in terms of amplitudes ${\cal
A}_{||}$ and ${\cal A}_{\perp}$. In particular, the \textit{unpolarized} angle-differential cross
section (averaged over the initial-photon polarizations and summed over the final-photon
polarizations) is just
\begin{equation} \label{eq2b}
\frac{d\sigma^{\rm unpol}}{d\Omega} = \frac12 \bigl(\big|{\cal A}_{||}\big|^2 + \big|{\cal A}_{\perp}\big|^2\bigr)\,.
\end{equation}

For the linearly polarized incident radiation, the angle-differential cross section is
\cite{roy:86}
\begin{align}\label{eq3}
\frac{d\sigma^{\rm lin}}{d\Omega} = &\
  \frac14 \big( 1 + \xi_{1i}\, \xi_{1f} \big) \bigl(\big|{\cal A}_{||}\big|^2 + \big|{\cal A}_{\perp}\big|^2\bigr)
 \nonumber \\
  &\
  + \frac14 \big( \xi_{1i} + \xi_{1f} \big)\, \bigl(\big|{\cal A}_{||}\big|^2 - \big|{\cal A}_{\perp}\big|^2\bigr)
  \,,
\end{align}
where $\xi_1 = \epsilon_{||}^2 - \epsilon_{\perp}^2$. In the particular case when the incident
radiation is fully polarized and the polarization of the scattered radiation is not observed, the
angle-differential cross section is defined only by the parallel amplitude,
\begin{align}\label{eq3b}
\frac{d\sigma^{\rm lin}}{d\Omega} = &\
  \big|{\cal A}_{||}\big|^2 \,.
\end{align}

In the present work we are specifically interested in the Stokes parameter $P_1$ of the scattered
radiation, which can be evaluated as
\begin{equation}\label{eq4}
   P_1 = \frac{{d} \sigma^{\rm lin}_{||}/ {d} \Omega - {d} \sigma^{\rm lin}_{\perp}/ {d} \Omega}
              {{d} \sigma^{\rm lin}_{||}/ {d} \Omega + {d} \sigma^{\rm lin}_{\perp}/ {d} \Omega} \, ,
\end{equation}
where ${d} \sigma^{\rm lin}_{||}/ {d} \Omega$ and ${d} \sigma^{\rm lin}_{\perp}/ {d} \Omega$ are
the differential cross sections for the scattering of radiation polarized parallel to and
perpendicular to the scattering plane, respectively. Using Eq.~(\ref{eq3}) and identifying the
Stokes parameter $P_1$ of the incoming light as $P_{1i} = \xi_{1i}$, we obtain
\begin{equation}\label{eq5}
   P_{1\!f} = \frac{(1+P_{1i})\big|{\cal A}_{||}\big|^2 - (1-P_{1i})\big|{\cal A}_{\perp}\big|^2}
                 {(1+P_{1i})\big|{\cal A}_{||}\big|^2 + (1-P_{1i})\big|{\cal A}_{\perp}\big|^2}\,.
\end{equation}
This expression connects the degree of polarization of the incident and the scattered radiation. It
shows, in particular, that the scattered radiation is always fully polarized ($P_{1\!f} = 1$) when
the initial radiation is fully polarized ($P_{1i} = 1$).

\section{Calculation of amplitudes}

In this section we discuss the calculation of the amplitudes ${\cal A}_{||}$ and ${\cal
A}_{\perp}$, which are the main building blocks for all observable quantities.

Let us first define the geometry of the process. We chose the $z$ axis of our Cartesian coordinate
system along the momentum of the incoming photon ${\bm k}_i$ and the $x$ axis along the
polarization vector of the incoming photon ${\bm \epsilon}_i$. With this choice of the coordinate
system, the angle-differential cross section depends on angles only through the polarization vector
of the outgoing photon ${\bm \epsilon}_f$ and $\hat{\bm k}_f \equiv {\bm k}_f/|{\bm k}_f|=
(\theta,\phi)$, where $\cos \theta = \hat{\bm k}_i \cdot \hat{\bm k}_f$. The plane spanned by
vectors ${\bm k}_i$ and vectors ${\bm k}_f$ is commonly referred to as the scattering plane and the
angle $\theta$ between them, as the scattering angle.

Our calculations of the transition amplitudes of the Rayleigh scattering are performed within the
independent particle approximation (IPA), in which the amplitudes of the scattering off individual
electrons of the atomic target are taken to be additive. From these shell-dependent amplitudes, the
total scattering amplitude is obtained as a sum
\begin{equation} \label{eq6aa}
{\cal M}_{fi} = \sum_{n} {\cal M}_{n} = \sum_{njl} {\cal M}_{njl}\,,
\end{equation}
where $n$, $l$, and $j$ are the quantum numbers of the one-electron states and the summation runs
over all occupied states. The individual contributions ${\cal M}_{n}$ with $n = 1$, 2, 3, etc. are
referred to as the contributions of the $K$, $L$, $M$, $\ldots$ shells. The IPA approximation is
known to work fairly well, as discussed in Refs.~\cite{kissel:80,roy:86,roy:99}. This conclusion
was also confirmed by numerical calculations of the electron-electron interaction corrections to
Rayleigh scattering recently performed for helium-like ions in Ref.~\cite{volotka:16}.

In the previous work \cite{surzhykov:15} we described our approach to the calculation of the
one-particle amplitudes ${\cal M}_{njl}$, which is based on a numerical solution of the Dirac
equation for an electron in the central field of the nucleus and the ``spectator'' electrons.
Specifically, we apply the power series expansion method for solving the radial Dirac equation with
a general scalar potential $V(r)$ for an arbitrary energy $E$. Using the asymptotic form of the
exact Dirac functions, we compute the regular and irregular solutions of the Dirac equation and
store them on a radial grid. Knowing the regular and irregular Dirac solutions, we construct the
Green function of the Dirac equation, which is then used for computing the second-order matrix
elements in Eq.~(\ref{eq1}). The numerical method is described in Ref.~\cite{yerokhin:11:fns}.

An important feature of the computation is a summation over the complete one-electron spectrum of
the Dirac equation and, as a consequence, an infinite summation over the multipoles of the photon
field $L$ (which is referred to a as the partial-wave (PW) expansion). In our calculations, we
extended the PW expansion up to $L_{\rm max} = 70$, which yielded converged results for the
amplitudes of the $K$, $L$, $M$, and, sometimes, $N$ shells in the expansion (\ref{eq6aa}). For the
higher-$n$ atomic shells, however, the convergence of the PW expansion becomes increasingly slower.
Fortunately, numerical contributions of the shells with $n > 4$ decrease quickly, except for the
forward scattering angles.

The contributions of  the higher-$n$ shells can be accounted for within the so-called form-factor
(FF) approximation, in which the photon scatters off a static charge distribution of individual
electrons as obtained from the Dirac-Fock equation. Pratt and co-workers \cite{kissel:80,kane:86}
reported a modified FF approximation that included some binding effects and estimates for the
imaginary part of the amplitudes. Following Refs.~\cite{kissel:80,kane:86}, we define the
approximate FF amplitudes by
\begin{align}
{\rm Re}\,{\cal A}_{||,n} &\ = - G_n(q) \, \cos\theta\,, \\
{\rm Re}\,{\cal A}_{\perp,n} &\ = - G_n(q)\,, \\
{\rm Im}\,{\cal A}_{||,n} &\ = \left( \sigma^{\rm PE}_n/\sigma^{\rm PE}_{n_0}\right)\,
    {\rm Im}\,{\cal A}_{||,n_0}\,,\label{12}\\
{\rm Im}\,{\cal A}_{\perp,n} &\ = \left( \sigma^{\rm PE}_n/\sigma^{\rm PE}_{n_0}\right)\,
    {\rm Im}\,{\cal A}_{\perp,n_0}\,,\label{13}
\end{align}
where $G_n(q)$ is the form-factor of the $n$ shell, $\sigma^{\rm PE}_n$ is the cross section for
the photoionization from the $n$th shell, and $n_0$ is an inner shell for which partial-wave
expansion results are available. The modified form-factor is defined as \cite{franz:36}
\begin{align}
G_n(q) = \int_0^{\infty} r^2dr\, \rho_n(r)\,\frac{\sin qr}{qr}\, \frac1{\vare_n - V_n(r)}\,,
\end{align}
where $q = 2\,\omega\sin (\theta/2)$ is the momentum transferred to the target atom by the
scattered photon, $\vare_n$ is the energy of an electron in the $n$th shell (including the rest
mass), $V_n(r)$ is the binding potential for electrons in the $n$th shell (including the nuclear
Coulomb field and the screening potential from the other shells), and $\rho_n(r)$ is the electron
charge density of the $n$th shell, normalized to the number of electrons in the shell, $
\int_0^{\infty} r^2dr\, \rho_n(r) = 2j_n+1\,. $

In our numerical calculations, we use the partial-wave expansion (PWE) method to compute the
amplitudes of the inner $n \leq n'$ shells (with $n' = 3$ or 4), and the modified FF approach to
approximate the amplitudes for the remaining outer shells. This approach was previously used in
calculations of Pratt and co-workers \cite{kissel:80,kane:86}.

Table~\ref{tab:1} compares results of the PWE calculation with those obtained in the modified FF
approximation, for the case of scattering off neutral lead ($Z = 82$) and the photon energy of
$\omega = 145$~keV. We observe that the FF approximation works very well for forward scattering
angles. For larger angles and higher scattering energies, however, its accuracy gradually
deteriorates. The use of the FF approximation is essential in order to obtain reliable results for
the forward scattering, since the contribution of the outer $n
> 4$ shells (for which no convergent PWE results are obtained) is significant in this case. For
larger angles, however, the numerical contribution of the outer shells is small and can often be
omitted.

\begin{figure}[t]
\centerline{
\resizebox{0.5\textwidth}{!}{%
  \includegraphics{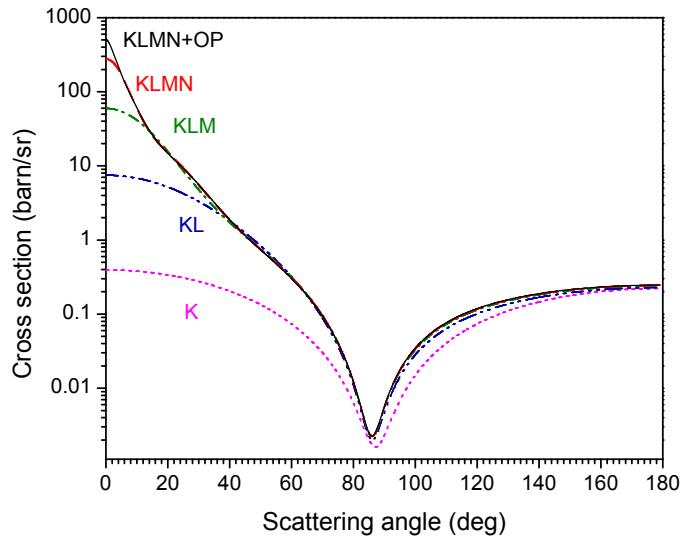}
}}
 \caption{Angle-differential cross section of the Rayleigh scattering of fully linearly polarized $145$~keV x-rays off the neutral
 lead atom (Pb, $Z = 82$). Solid line (black) shows results of the full calculation, with the PWE treatment of the
$KLMN$ shells and the FF treatment of the $OP$ shells. The other lines present results of the PWE calculations with inclusion of
the $KLMN$ shells (dashed line, red), $KLM$ shells (dashed-dotted line, green), $KL$ shells (dashed-dot-dotted line, blue),
and $K$ shells (dotted line, magenta).
 \label{fig:1}}
\end{figure}

Figure~\ref{fig:1} displays the convergence of calculated results with respect to the number of
shells taken into account, for the angle-differential cross section $d\sigma(\theta,\phi =
0)/d\Omega$ for the completely linearly-polarized incident light with energy $\omega = 145$~keV and
neutral lead target. We observe that the contribution of shells with $n>4$ is noticeable only in
the very forward direction. We performed two variants of the full (PWE+FF) calculation, evaluating
the contribution of the $n = 4$ shell by the PWE and FF methods. Both computations yielded results
that are visually indistinguishable on the picture.

\begin{table*}
\caption{Rayleigh-scattering amplitudes ${\cal A}_{||}$ and ${\cal A}_{\perp}$
for Pb ($Z = 82$) at photon energy $\omega = 145$~keV, in relativistic units.
``PWE'' labels results obtained within the partial wave expansion, ``FF''
denotes results obtained within the modified formfactor approximation. $n$ is the shell number ($n = 1$ is $K$ shell,
$n = 2$, $L$ shell, etc.). The imaginary part of the FF amplitudes is scaled to match the corresponding PWE
amplitude for the $n=3$ shell, according to Eqs.~(\ref{12}) and (\ref{13}).
\label{tab:1} }
\begin{ruledtabular}
\begin{tabular}{rcrrrrrrrr}
\multicolumn{1}{r}{$\theta$}
    & \multicolumn{1}{c}{$n$}
        & \multicolumn{2}{c}{${\cal A}_{||}$(PWE)}
        & \multicolumn{2}{c}{${\cal A}_{||}$(FF)}
        & \multicolumn{2}{c}{${\cal A}_{\perp}$(PWE)}
        & \multicolumn{2}{c}{${\cal A}_{\perp}$(FF)} \\
     &       & \multicolumn{1}{c}{Re}        & \multicolumn{1}{c}{Im}
             & \multicolumn{1}{c}{Re}        & \multicolumn{1}{c}{Im}
             & \multicolumn{1}{c}{Re}        & \multicolumn{1}{c}{Im}
             & \multicolumn{1}{c}{Re}        & \multicolumn{1}{c}{Im} \\
\hline\\[-5pt]
%
%
  0  &    1  &   $-$1.9196  &   1.1357  &    $-$1.7238  &   1.0970  &    $-$1.9196  &   1.1357  &    $-$1.7238  &   1.0970   \\
     &    2  &   $-$7.6961  &   0.2087  &    $-$7.6736  &   0.2047  &    $-$7.6961  &   0.2087  &    $-$7.6736  &   0.2047  \\
     &    3  &  $-$17.7637  &   0.0480  &   $-$17.7633  &   0.0480  &   $-$17.7637  &   0.0480  &   $-$17.7633  &   0.0480  \\
     &    4  &  $-$31.8873  &   0.0118  &   $-$31.8849  &   0.0122  &   $-$31.8873  &   0.0118  &   $-$31.8849  &   0.0122  \\
     &    5  &            &             &   $-$17.9860  &   0.0025  &             &             &   $-$17.9860  &   0.0025  \\
     &    6  &            &             &    $-$3.9995  &   0.0002  &             &             &    $-$3.9995  &   0.0002  \\
  \hline \\[-5pt]
 30  &    1  &   $-$1.6287  &   0.9048  &    $-$1.4514  &   0.7763  &    $-$1.8748  &   1.1001  &    $-$1.6760  &   1.0242  \\
     &    2  &   $-$4.7296  &   0.1494  &    $-$4.8036  &   0.1448  &    $-$5.5312  &   0.1956  &    $-$5.5467  &   0.1911  \\
     &    3  &   $-$1.2674  &   0.0339  &    $-$1.3791  &   0.0339  &    $-$1.4904  &   0.0448  &    $-$1.5925  &   0.0448  \\
     &    4  &   $-$0.5964  &   0.0084  &    $-$0.6185  &   0.0086  &    $-$0.6945  &   0.0110  &    $-$0.7142  &   0.0114  \\
     &    5  &            &             &    $-$0.0867  &   0.0018  &             &             &    $-$0.1001  &   0.0023  \\
     &    6  &            &             &    $-$0.0016  &   0.0001  &             &             &    $-$0.0019  &   0.0002  \\
  \hline \\[-5pt]
 60  &    1  &   $-$0.8851  &   0.3701  &    $-$0.7775  &   0.1806  &    $-$1.7660  &   1.0142  &    $-$1.5550  &   0.8673  \\
     &    2  &   $-$1.0626  &   0.0369  &    $-$1.2121  &   0.0337  &    $-$2.3606  &   0.1669  &    $-$2.4242  &   0.1618  \\
     &    3  &      0.0110  &   0.0079  &    $-$0.0170  &   0.0079  &    $-$0.0249  &   0.0379  &    $-$0.0340  &   0.0379  \\
     &    4  &      0.0261  &   0.0020  &       0.0206  &   0.0020  &       0.0406  &   0.0093  &       0.0411  &   0.0096  \\
     &    5  &              &           &       0.0032  &   0.0004  &               &           &       0.0064  &   0.0020  \\
     &    6  &              &           &       0.0001  &   0.0000  &               &           &       0.0001  &   0.0002  \\
  \hline \\[-5pt]
120  &    1  &    0.8031  &  $-$0.5380  &     0.6421  &  $-$0.4462  &    $-$1.5533  &   0.8397  &    $-$1.2843  &   0.6115  \\
     &    2  &    0.1391  &  $-$0.0862  &     0.1440  &  $-$0.0832  &    $-$0.1876  &   0.1192  &    $-$0.2880  &   0.1141  \\
     &    3  &    0.0696  &  $-$0.0195  &     0.0731  &  $-$0.0195  &    $-$0.1198  &   0.0267  &    $-$0.1462  &   0.0267  \\
     &    4  &    0.0194  &  $-$0.0048  &     0.0205  &  $-$0.0050  &    $-$0.0341  &   0.0066  &    $-$0.0410  &   0.0068  \\
     &    5  &            &             &     0.0041  &  $-$0.0010  &               &           &    $-$0.0081  &   0.0014  \\
     &    6  &            &             &     0.0002  &  $-$0.0001  &               &           &    $-$0.0004  &   0.0001  \\
\end{tabular}
\end{ruledtabular}
\end{table*}

\section{Determination of the polarization purity of the initial photon}

\begin{figure}[!htb]
\centerline{
\resizebox{0.5\textwidth}{!}{%
  \includegraphics{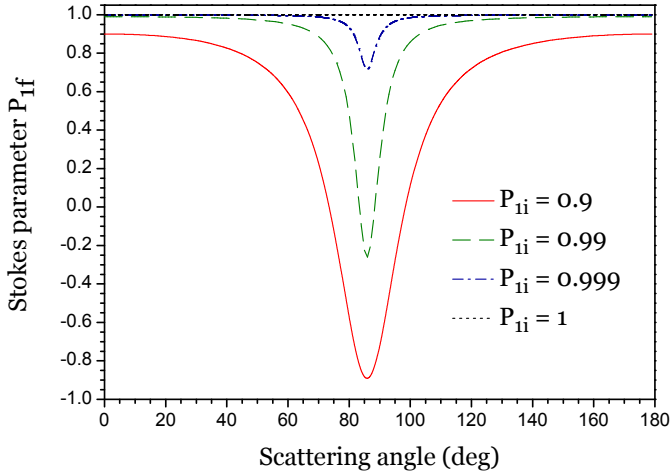}
}}
 \caption{The Stokes parameter of the scattered radiation $P_{1\!f}$ as a function of the scattering angle $\theta$,
 for different degrees of linear polarization of the incoming radiation $P_{1i}$, for scattering of the
 $145$~keV x-rays off the neutral lead atom (Pb, $Z = 82$).
 \label{fig:P1f}}
\end{figure}

We now consider a correlation between the degree of linear polarization of the incoming and
scattered x-rays. Already from Eq.~(\ref{eq2}) it can be seen that if the incident radiation is
polarized parallel or perpendicular to the scattering plane, the scattered radiation will have
exactly the same polarization. A general relation between the Stokes parameters $P_1$ of the
incident and scattered radiation is given by Eq.~(\ref{eq5}). Using the numerical results for the
transition amplitudes obtained in the previous section, we can deduce results also for the Stokes
parameters.

Fig.~\ref{fig:P1f} presents a typical example of the angular dependence of the Stokes parameter of
the scattered radiation $P_{1\!f}$ for different values of the Stokes parameter of the incident
radiation $P_{1i}$. We observe that in the region of scattering angles $\theta \approx 90^{\circ}$
the degree of polarization of the scattered radiation depends strongly on the degree of
polarization of the initial radiation. If, for example, the incident radiation is depolarized by
just $0.1\%$, the depolarization of the scattered radiation might reach up to $20\%$.

From Fig.~\ref{fig:1} we conclude that the vicinity of scattering angles $\theta \approx
90^{\circ}$ is the region of the suppressed photon emission and, according to Eq.~(\ref{eq3b}), of
small values of the parallel amplitude $A_{||}$. (We recall that in the formfactor approximation,
$A_{||}\propto \cos^2 \theta$). For the angles $\theta \approx 90^{\circ}$ and the incident
radiation of high polarization purity ($1-P_{1\!i} \ll 1$), the numerator and the denominator of
Eq.~(\ref{eq5}) are composed of two small quantities, $(1+P_{1\!i})|A_{||}|^2$ and
$(1-P_{1\!i})|A_{\perp}|^2$, so that even small variations of the amplitudes and the initial
polarization may lead to large changes of the polarization of the scattered radiation.

This effect can be used as a method for an accurate determination of small depolarization of
initial photon $\delta_{i} \equiv 1- P_{1i}$, by measuring the Stokes parameter of the scattered
radiation $P_{1\!f}$ at scattering angles $\theta$ around $90^{\circ}$. We note that despite the
strong suppression of the cross section at $\theta \approx 90^{\circ}$, this angle region is still
accessible for experimental investigations \cite{blumenhagen:16}, due to great luminosity of
synhrotron radiation on modern facilities.

This remarkable effect can be understood more easily if we rewrite Eq.~(\ref{eq5}) in terms of the
depolarization of the initial radiation $\delta_{i}$ as
\begin{align}\label{eq10}
P_{1\!f} = 1 - \frac{2}{1 + \frac{\displaystyle 2-\delta_{i}}{\displaystyle \delta_{i}}\,R(E,\theta)}\,,
\end{align}
where $R$ is the ratio of the squares of the parallel and perpendicular amplitudes,
\begin{align}\label{eq11}
R(E,\theta) = \frac{\big| A_{||} \big|^2}{\big| A_{\perp} \big|^2}\,.
\end{align}
We note that within the (standard) form-factor approximation, $R(E,\theta)$ does not depend on
energy and is just $R(E,\theta) = R_{\rm FF}(\theta) = \cos^2\theta$, so that Eq.~(\ref{eq10})
becomes an universal function, without any dependence on the nuclear charge of the target or the
energy of the incident radiation.

In order to study the sensitivity of $P_{1\!f}$ on $\delta_{i}$, we differentiate Eq.~(\ref{eq10}),
obtaining
\begin{align}\label{eq12}
\frac{\partial P_{1\!f}}{\partial \delta_{i}}
= -\frac{4 R(E,\theta)}{\big[R(E,\theta)(2-\delta_{i})+\delta_{i}\big]^2}\,.
\end{align}
The derivative behaves as $-1/(2\delta_i)$ for $R\approx \delta_{i}/2 \ll 1$ and as $-1/R$ for
$\delta_{i} \ll R \ll 1$. We conclude that by measuring the Stokes parameter of the outgoing
radiation $P_{1\!f}$ in the region of small values of $R(E,\theta)$, we achieve an enhanced
sensitivity of our measurement to small depolarizations of the incoming radiation.

\begin{figure}[!htb]
\centerline{
\resizebox{0.5\textwidth}{!}{%
  \includegraphics{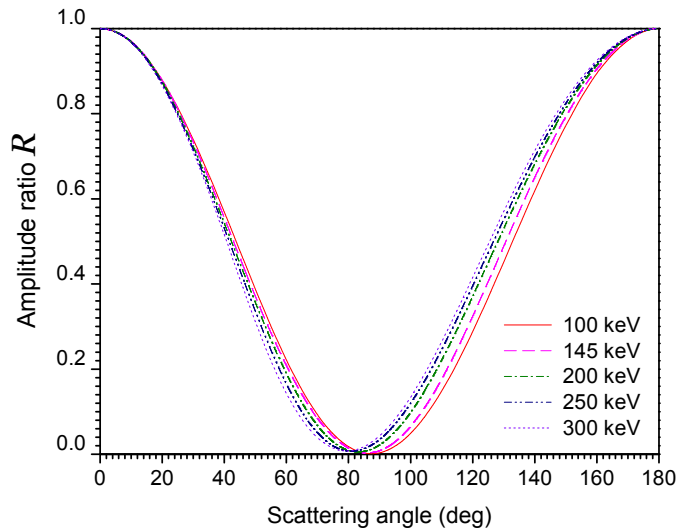}
}}
 \caption{The ratio of the squares of the parallel and perpendicular amplitudes $R(E,\theta)$ as
a function of the scattering angle $\theta$,
for scattering of the x-rays off neutral lead atom (Pb, $Z = 82$). \label{fig:Ra}
}
\end{figure}

\begin{figure}[!htb]
\centerline{
\resizebox{0.5\textwidth}{!}{%
  \includegraphics{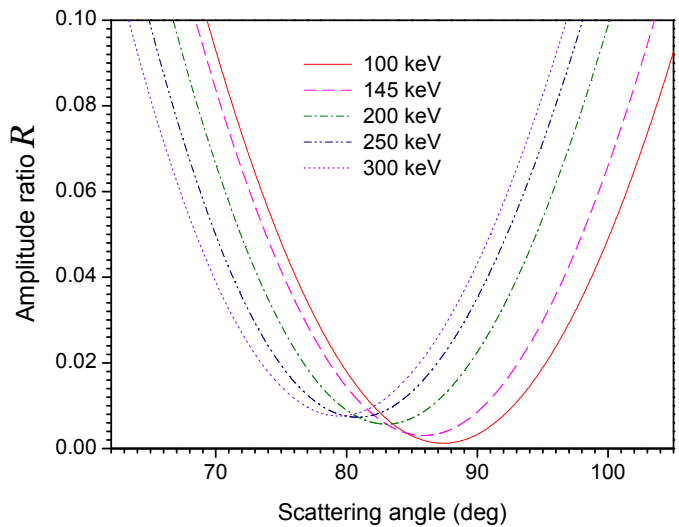}
}}
 \caption{A fragment of Fig.~\ref{fig:Ra}, for the angle region relevant for the determination
of the polarization purity of the initial radiation.  \label{fig:Rb}
}
\end{figure}

The angular dependence of the function $R(E,\theta)$ for different collision energies is plotted on
Figs.~\ref{fig:Ra} and \ref{fig:Rb}, for scattering off the neutral lead target. The plotted
results were obtained within the PWE approach with including shells with $n\le 4$ for energies 100
and 145~keV and $n\le 3$ for higher energies.

Contrary to the FF approximation, in which the function $R$ vanishes at $\theta = 90^{\circ}$, the
PWE calculation shows that the minimal value of $R$ is always above zero. The minimal value
gradually increases with the increase of the collision energy. The \textit{position} of the minimum
also shifts from the FF value $\theta = 90^{\circ}$ toward smaller angles; the higher the energy,
the larger the shift. We observe that for a wide range of the collision energies, the angle region
of $\theta = 60^{\circ}$-$100^{\circ}$ corresponds to values of $R \,\lapprox \,0.1$, thus yielding
an enhancement of sensitivity in determination of $\delta_{i}$ from one to two orders of magnitude.


The method of determining $\delta_{i}$ from Eq.~(\ref{eq10}) discussed so far depends on
theoretical results for the function $R(E,\theta)$, which need to be obtained reliably and with a
controlled accuracy. In Table~\ref{tab:R} we present results of our calculations of $R(E,\theta)$
for the neutral lead target and different angles and scattering energies. In our calculation of
$R(E,\theta)$, we found that the inclusion of the FF approximation for the higher-$n$ shells does
not improve the convergence of the calculated results for the scattering angles we are presently
interested, so we did not include it. The uncertainty ascribed to the numerical values of
$R(E,\theta)$ reflects our estimation of the error caused by an incomplete treatment of the outer
electron shells.

We neglect uncertainty due to smaller effects not included into calculation, even though they might
be enhanced in the region of $\theta \approx 90^{\circ}$. Among these effects, the largest one is
probably the residual electron correlation, studied (to first order in $1/Z$ and for He-like ions)
in Ref.~\cite{volotka:16}. The results of Ref.~\cite{volotka:16} suggest that an enhancement around
90$^{\circ}$ exists but the change of the cross section is still very small.

One can avoid the need of theoretical predictions for $R(E,\theta)$ by measuring an independent
observable that depends on $\delta_i$ and $R(E,\theta)$. We propose to use for this purpose the
ratio of the angle-differential cross sections measured \textit{within} and \textit{perpendicular}
to the plane of polarization of the incident radiation. This ratio can be evaluated as
\begin{align}\label{eq14}
\sigma_R \equiv \frac{d\sigma({\bm \epsilon}_{i||})/(d\Omega)}{d\sigma({\bm \epsilon}_{i\perp})/(d\Omega)}
 = \frac{1 + \frac{\displaystyle 2-\delta_{i}}{\displaystyle \delta_{i}}\,R(E,\theta) }
    {R(E,\theta)+\frac{\displaystyle 2-\delta_{i}}{\displaystyle \delta_{i}}}\,,
\end{align}
where we assumed that the polarization of the scattered photons remains unobserved.

Therefore, by measuring $P_{1\!f}$ and $\sigma_R$ for the same scattering energy $E$ and scattering
angle $\theta$, we can determine the degree of polarization of the initial radiation $\delta_{i}$
from Eqs.~(\ref{eq10}) and (\ref{eq14}), without any further theoretical input.

\section{Conclusion}

We have studied the polarization properties of x-ray radiation elastically scattered off a
closed-shell atomic target. We have shown, in particular, that the degree of polarization of the
scattered radiation is very sensitive to small deviations of the incident radiation from the 100\%
linear polarization, especially in the region of scattering angles $\theta = 80$--$90^{\circ}$.
This effect can be used as a precision tool for the diagnostics of the polarization purity of the
synchrotron radiation. Two variants of the diagnostics method have been put forward. The first one
requires the ratio of the squared amplitudes $R(\theta,E)$ to be taken from theoretical
calculations, whereas the second one determines $R(\theta,E)$ from an additional measurement of the
ratio of the scattering cross section observed within and perpendicular to the polarization plane
of the incident radiation.

\begin{acknowledgments}
   V.A.Y. and A.V. acknowledge support by the Russian Foundation for Basic Research Grant No.
   16-02-00538-a. V.A.Y. acknowledges also support from
   the Ministry of Education and Science of the Russian Federation Grants No. 3.5397.2017/6.7 and 3.1463.2017/4.6.
\end{acknowledgments}

\begin{table}
\caption{The ratio of the squares of the parallel and perpendicular amplitudes
$R(E,\theta)$ for Rayleigh scattering off the neutral lead atom. ``FF'' denotes the (standard) form-factor
approximation, with $R_{\rm FF}(\theta) = \cos^2\theta$.
\label{tab:R} }
\begin{ruledtabular}
\begin{tabular}{ccd}
\multicolumn{1}{r}{$\theta$ (deg)}
    & \multicolumn{1}{c}{$E$ (keV)}
        & \multicolumn{1}{c}{$R$ }
        \\\hline\\[-5pt]
30&	FF  &	0.7500       \\
  &	100 &   0.7394\,(3)  \\
  &	145 &   0.7335\,(3)  \\  
  &	200	&	0.7224\,(4)	 \\  
  &	250	&	0.7148\,(10) \\  
  &	300	&	0.7095\,(15) \\[5pt]
60& FF  &   0.2500      \\
  &	100	&	0.2212\,(4)  \\
  & 145	&	0.2077\,(6)	 \\ 
  &	200	&	0.1888\,(7)	 \\ 
  &	250	&	0.1646\,(9)  \\ 
  &	300	&	0.1425\,(15) \\[5pt]
70& FF  &   0.1170 \\
  &	100	&	0.0926\,(4) \\
  & 145	&	0.0837\,(4)  \\ 
  &	200	&	0.0663\,(6)  \\ 
  &	250	&	0.0497\,(7)  \\ 
  &	300	&	0.0386\,(15) \\[5pt]
80& FF  &   0.0302 \\
  &	100	&	0.0176\,(2) \\
  & 145	&	0.0143\,(1)  \\ 
  &	200	&	0.0088\,(2)  \\ 
  &	250	&	0.0074\,(2)  \\ 
  &	300	&	0.0077\,(3)  \\[5pt]
90&	FF  &   0.0000       \\
  &	100	&	0.0030\,(1)  \\
  &	145	&	0.0085\,(1)	 \\  
  &	200	&	0.0225\,(2)  \\  
  &	250	&	0.0352\,(8)	 \\  
  &	300	&	0.0436\,(18) \\
\end{tabular}
\end{ruledtabular}
\end{table}

\end{document}